\documentclass[review]{elsarticle}
\usepackage{amssymb}
\usepackage{amsmath}
\usepackage{graphicx}
\usepackage{float}
\usepackage{xcolor}
\usepackage{hyperref}
\usepackage{subfig}

\journal{Physica A: Statistical Mechanics and its Applications}

\begin{document}

\begin{frontmatter}

\title{Modularity and Projection of Bipartite Networks}

\author{Rudy Arthur}
\address{Department of Computer Science, University of Exeter, North Park Road,Exeter,UK,EX4 4QF}




\begin{abstract}
This paper investigates community detection by modularity maximisation on bipartite networks. In particular we are interested in how the operation of projection, using one node set of the bipartite network to infer connections between nodes in the other set, interacts with community detection. We first define a notion of modularity appropriate for a projected bipartite network and outline an algorithm for maximising it in order to partition the network. Using both real and synthetic networks we compare the communities found by five different algorithms, where each algorithm maximises a different modularity function and sees different aspects of the bipartite structure. Based on these results we suggest a simple heuristic for finding communities in bipartite networks.
\end{abstract}

\begin{keyword}
Networks, Community Detection, Bipartite, Modularity
\end{keyword}

\end{frontmatter}


\section{Introduction}

Bipartite graphs are ubiquitous in network theory. To cite a few of the better known examples; networks of movies and actors \cite{barabasi:1999}, papers and co-authors \cite{newman:2001}, words and documents \cite{Dhillon:2001:CDW:502512.502550}, disorders and genes \cite{Goh8685}, consumers and purchases \cite{Matsuo:2009} can all be represented as bipartite networks. A bipartite network consists of two sets of nodes of different type e.g. actors in one set and movies in the other, and connections only between nodes in different sets. These conditions mean that many of the metrics and algorithms developed for unipartite/one-node-type networks do not work `out of the box'. 

The two main approaches to dealing with bipartite structure have been
\begin{enumerate}
    \item Modifying unipartite algorithms to account for the bipartite structure.
    \item Projecting the bipartite graph onto one of its node sets.
\end{enumerate}
We will be interested in the second approach for the particular problem of community detection. The first approach would seem to side-step many issues but it is still valuable to study the projection problem. Many useful unipartite metrics or algorithms cannot be extended to the bipartite case and often, even when they can, the implementation is only available in specialist software and not in general purpose network analysis tool-kits like Gephi\cite{gephi} or NetworkX\cite{networkx}. The idea of projection is also intuitively reasonable. The network of actors linked based on shared credits for example, has been a benchmark of network analysis since its early days, however the optimal way to weight each of these links is much debated \cite{newman2001scientific, PhysRevE.76.046115, LI20133248}. Finally, we may not always have access to the bipartite network. Our network may be better described by a bipartite network but the bipartite structure may be hidden. For example, connections in an online social network can be made at parties, hobbies or at work but the person/event graph is not observable, we only observe its `projection' through the friend/follower connection on the social network. 

It is therefore important to know what kind of information is lost and what is preserved when projecting a bipartite network. Certainly there are different bipartite network motifs which lead to the same projected unipartite graph \cite{PhysRevE.78.016108}. However analysis of network's eigenvalue spectrum suggests that much of the information from the original graph is preserved, especially if both projections are available \cite{Everett2013TheDA, melamed2014community}. 

In this work we address the problem of community detection and bipartite projection. The community detection problem \cite{Girvan2002Community, FORTUNATO201075} asks for a partition of the nodes into disjoint sets (though see e.g. \cite{PalEtAl05} where overlapping sets are allowed) such that nodes in the same set are more densely connected to each other than to nodes not in the set. Though many approaches to this problem exist e.g. the stochastic block model \cite{holland1983stochastic}, betweenness centrality \cite{newman2001scientific}, spectral methods \cite{newman2013spectral}, the most popular approaches, and the ones we will focus on, are based on optimising modularity \cite{newman2004fast}. The questions we will try to answer in this work are the following. Does projecting destroy community structure that is present in the bipartite graph or induce spurious community structure that is not present in the unprojected graph? If so, what kinds of bipartite graphs suffer more or less from these problems? And crucially, what should one do in practice for detecting communities in a bipartite graph?

{\color{black} There are two parts to our work. The first part is formulating a definition of modularity that is appropriate for projections of bipartite networks and describing a community detection algorithm based on this definition. The second part is a comparison between the different modularity functions which can be used to detect communities in bipartite networks. The aim of this is to examine the interaction between projection and community detection. We will identify the conditions a network should satisfy so that communities detected after projection reflect the true structure of the network and are not distorted by the projection operation. Based on this we make suggestions for how to approach community detection on bipartite graphs in practice.

In section \ref{sec:mod} we review the definition modularity, how it has been modified for bipartite graphs and introduce our own definition suitable for working with projected bipartite graphs. In section \ref{sec:Synthetic} we describe how to construct bipartite graphs with specified community structure and specified degree distributions, similar to the LFR benchmarks \cite{lancichinetti2008benchmark}. Section \ref{sec:alg} outlines the algorithms we will use to detect community structure. Section \ref{sec:sim} describes the metrics we use to measure partition similarity, we use three metrics which respond to different error modes, giving a richer picture of the quality of the partitions found by the community detection algorithms. In section \ref{sec:syn} we attempt to address the questions posed above using the synthetic networks: namely when projection affects the community structure that can be detected by modularity based methods}. In section \ref{sec:real} we see how the same approaches work on real bipartite networks before concluding in section \ref{sec:conclusion} and suggesting some practical heuristics for community detection.

\section{Modularity and Communities}\label{sec:mod}

Modularity makes formal our intuitive sense that a network's communities should have more internal links than external ones. Consider a partition of the nodes, $c(i)$, mapping each node label, $i$, to a community label. Modularity is the difference between the proportion of within community nodes in the original network and the proportion that would be expected using the same partition and re-wiring the network while keeping the degrees of all the nodes fixed.
\begin{equation}\label{eqn:mod}
Q = \frac{1}{2E} \sum_{ij} \left( A_{ij} - \frac{k_i k_j
}{2E} \right) \delta( c(i), c(j) )
\end{equation}
Here $A$ is the adjacency matrix of the network, $k_i = \sum_j A_{ij}$ is the degree of node $i$ and $2E = \sum_{i} k_i$ is twice the number of edges in the network. Different null models are possible and a more general modularity definition is
\begin{equation}
Q = \frac{1}{2E} \sum_{ij} \left( A_{ij} - \lambda P_{ij} \right) \delta( c(i), c(j) )
\end{equation}
where $P_{ij}$ is the probability of nodes $i$ and $j$ being connected in the null model. $\lambda$ is the so-called resolution parameter \cite{Fortunato36, lambiotte2008laplacian} which we will set to 1 in the following, but which can be used to force the optimal modularity partition to contain more or fewer communities.

For a bipartite network, the constraint that no connections can exist between nodes of the same type has inspired several different definitions of modularity. The most common, and the one we will use in this paper, is due to Barber \cite{barber2007modularity} who defined the modularity of a bipartite graph as
\begin{equation}\label{eqn:bmod}
Q_B = \frac{1}{F} \sum_{ij} \left( B_{ij} - \frac{q_i d_j
}{F} \right) \delta( c(i), c(j) )
\end{equation}
where $B$ is the $N \times M$ bipartite adjacency matrix, $q_i = \sum_{j} B_{ij}$ and $d_j = \sum_i B_{ij}$ and $F = \sum_j d_j = \sum_i q_i$. Other well known bipartite modularities are due to Guimera et. al. \cite{guimera2004modularity} and Murata \cite{murata2009detecting}, though we will not consider them here.

\subsection{Modularity for Projected Bipartite Networks}

We will project the bipartite graph onto one of its node types using a simple weighted projection.
\begin{equation}
A_{ij} = \sum_m B_{im} B_{jm}
\end{equation}
defines the adjacency matrix of the bipartite network after projection onto the $N$ `bottom' nodes. In what follows we will always refer to the nodes we have projected onto, e.g. actors, as bottom nodes and the nodes we have projected out, e.g. movies, as top nodes. The standard definition of modularity uses a null model based on the idea of rewiring connections while keeping the degrees of each node fixed. However when working with a projected bipartite graph we know that connections between nodes are induced by shared neighbours in the original graph. We propose a new definition of modularity that accounts for this fact.

As usual, we define modularity as the difference between two ratios. We measure the proportion of in-community edges in the projected graph using
\begin{equation}
    \frac{1}{2E} \sum_{ij} A_{ij} \delta( c(i), c(j) )
\end{equation}
where $A$ is the projected adjacency matrix. Our new null model is to randomly rewire edges \textit{in the original bipartite graph} and compute the projection. The probability of having a link between $i$ and $j$ in the projected, randomly rewired network is the probability of a link between $i$ and $m$ times the probability of having a link between $j$ and $m$, summed over all $m$
\begin{equation}
    \frac{1}{2E} \sum_{ijm} \frac{q_i q_j d_m^2}{F^2} \delta( c(i), c(j) ) = \sum_{ij} \frac{q_i q_j}{F^2} \delta( c(i), c(j) )
\end{equation}
where we used $2E = \sum_m d_m^2$. We define projected modularity as
\begin{equation}\label{eqn:pmod}
Q_P = \sum_{ij} \left( \frac{A_{ij}}{2E}  -  \frac{q_i q_j}{F^2} \right) \delta( c(i), c(j) )
\end{equation}

We now have two ways to measure the modularity of communities on a {\color{black} projected} bipartite graph: $Q$ where the null model rewires the projected graph or $Q_P$ where the null model rewires the original graph, then projects. This difference is crucial when dealing with high degree nodes. A node with $n$ edges in the bipartite graph induces $\sim n^2$ edges in the projected graph, thus the degrees $q_i$ and $k_i$ can be quite different for large $n$. Since the formation of cliques is one of the main ways projection hides network structure \cite{PhysRevE.78.016108}, using $Q_P$ should help us, and our algorithms, to `see' more of the network's structure.

\section{Synthetic Networks}\label{sec:Synthetic}

There is an extensive literature on how the degree distribution of the top and bottom nodes affects the degree distribution of the projected network \cite{PhysRevE.64.026118,NACHER20114636,MUKHERJEE20113602,vasques2018degree}. Some exact results are available for e.g. exponential and Poisson distributions but for the most interesting, and important, case of power law distributions we have to rely on stochastic simulations. If both the top and bottom node's degree distributions are power laws, $P(d) \sim d^{-\mu_2}$ and $P(q) \sim k^{-\mu_1}$, with exponents $\mu_2, \mu_1$, Vasques et. al. \cite{vasques2018degree} show that when the bottom nodes have a heavier-tailed degree distribution, $\mu_2 > \mu_1$, the degree distribution of the projected graph follows the degree distribution of the bottom node set. In the alternative case, $\mu_2 \leq \mu_1$ the degree distribution of the projected graph is power-law like (with exponent roughly $\mu_1-1$) but flattened, due to large cliques forming.

We generate synthetic bipartite networks with known degree distributions following \cite{vasques2018degree}. For each of the $N$ top nodes and $M$ bottom nodes we assign a degree value from the chosen distributions to give us top and bottom degree sequences $d_j$ and $q_i$. We check the equality $\sum_i^N q_i = \sum_j^M d_j$ and discard a random node from the set with the largest sum until it is satisfied. Finally we join the stubs at random. In the case where the average degrees in each node set are different we will end up with unbalanced graphs, i.e. $N \neq M$, see \cite{vasques2018degree} for details. To build in a target community structure we construct $C$ bipartite networks, with the same top and bottom distributions and starting $N$ and $M$. We then cut a fraction $p$ of the links at random and join the stubs without regard to the community structure. 

$p$ is the amount of noise in the network, similar to the LFR parameter $\mu$ \cite{lancichinetti2008benchmark}. For $p=0$ we have $C$ disconnected components while for $p=1$ we have no preference for the original communities. As $p$ increases the original community structure becomes undetectable \cite{PhysRevLett.101.078701,PhysRevLett.108.188701,PhysRevLett.107.065701}. At this point community detection algorithms will still work, in the sense of detecting sets of densely inter-connected nodes, but these communities will not correspond to the target structure. This can be a point of confusion, there are partitions of random graphs which have high modularity \cite{guimera2004modularity,reichardt2006statistical} even though there is no `real' community structure. In the following we will look at both modularity and agreement with the target structure as measures of success, remembering that these are not equivalent and can even be opposed.

\section{Algorithms}\label{sec:alg}

We will detect communities using both the bipartite and projected networks. For the bipartite network we use the following 3 approaches
\begin{enumerate}
    \item \textbf{Naive}. We run the Louvain algorithm \cite{blondel2008fast} on the bipartite graph, ignoring the constraints imposed by the bipartite structure and optimising the standard modularity \ref{eqn:mod}.
    
    \item \textbf{BiLouvain}. We run the modified Louvain algorithm described in \cite{zhou2018novel} on the bipartite graph optimising Barber Modularity \ref{eqn:bmod}.
    
    \item \textbf{Dual Projection}. Following \cite{melamed2014community} we run the Louvain algorithm on the top and bottom projections separately, optimising $Q_P$, and then use agglomerative clustering \cite{Girvan2002Community} to join the top and bottom communities in order to maximise Barber modularity, $Q_B$. 
\end{enumerate}
{\color{black} The Dual Projection approach usually optimises $Q$ at both stages. We have two choices for optimisation in the projection steps: $Q$ or $Q_P$ and two choices in the agglomerative clustering step $Q$ or $Q_B$. We chose to optimise $Q_P$ during the projection step and $Q_B$ during the agglomeration step so that the algorithm can `see' the bipartite structure as much as possible.}

For the projected network we will use 
\begin{enumerate}
\setcounter{enumi}{3}
    \item \textbf{Standard}. We run the Louvain algorithm \cite{blondel2008fast} on the projected graph optimising the standard modularity \ref{eqn:mod}.
    
    \item \textbf{Projected}. We run the Louvain algorithm \cite{blondel2008fast} on the projected graph optimising the projected modularity \ref{eqn:pmod}.
\end{enumerate}
For the projected network we will also compare the partitions found by the bipartite algorithms with the top nodes discarded.

The Louvain algorithm is a widely used and efficient method for producing near optimal community partitions. To implement the Louvain method using $Q_P$ we need two results. 
\begin{enumerate}
    \item The formula for the gain in modularity from adding a node to a community. To compute this we rewrite modularity \ref{eqn:pmod}
    \begin{equation}
Q_P = \sum_{c \in C} \left( \frac{\Sigma_{in}^{c}}{2E}  - \frac{ \left( \Sigma_{bitot}^{c} \right)^2}{F^2} \right) 
\end{equation}
where $\Sigma_{in}^{c} = \sum_{ij \in c} A_{ij}$ and $\Sigma_{bitot}^{c} = \sum_{j \in c} q_j$. The gain in modularity from adding a node $n$ that was in its own community into a new community $c$ is
\begin{equation}
    \left( \frac{2 k_n^{c}}{2E}  - \frac{  2 \Sigma_{bitot}^{c} q_n }{F^2}  \right) 
\end{equation}
where $k_n^{c} = \sum_{j \in c} A_{nj}$.

\item The relationship between the original and induced graphs. A community assignment on the bottom nodes creates an induced bipartite graph
\begin{align}
B'_{c m} = \sum_{i \in c} B_{im}     
\end{align}
this can then be projected
\begin{align}
\tilde{A}_{c d} &= \sum_{m} B'_{c m} B'_{d m} \\ \nonumber
&= \sum_m \sum_{i \in c} B_{i m} \sum_{j \in d}  B_{j m}
\end{align}
The same community assignment creates an induced graph on the projected graph
\begin{align}
A'_{c d} &= \sum_{i \in c, j \in d} A_{ij} \\ \nonumber
&= \sum_{i \in c, j \in d} \sum_m B_{i m} B_{j m} \\ \nonumber
&= \sum_m \sum_{i \in c} B_{i m} \sum_{j \in d}  B_{j m} = \tilde{A}_{c d}
\end{align}
That is, the operations of creating the induced graph and projecting the bipartite graph commute. 
\end{enumerate}
We modified the implementation of Blondel\footnote{https://perso.uclouvain.be/vincent.blondel/research/louvain.html} to compute partitions using this definition of modularity, as well as to implement the bi-Louvain method described in \cite{zhou2018novel}.

\section{Similarity Metrics for Communities}\label{sec:sim}

{\color{black} The partitions found by clustering algorithms can be wrong in two main ways: they can detect spurious sub-structure in the target communities (splitting) or they can fail to distinguish distinct communities (joining). Commonly used similarity metrics like the Rand Index or Mutual Information do not tell us which kind of error is occurring. }

To account for this we use the homogeneity, $H$, and completeness, $C$, scores of Rosenberg and Hirschberg \cite{Rosenberg:2007}. Denoting the target communities as classes and the detected communities as clusters: a \textit{homogeneous} clustering is one where all of the clusters contain only data points which are members of a single class. For example, the class labels $(0,0,1,1)$ are totally homogeneous, $H = 1$, with the cluster labels $(1,1,0,0)$, as well as with $(0,0,1,2)$. However, for the cluster labels $(0,1,1,1)$ we have $H < 1$ since the 1 cluster contains a member of the 0 class. Completeness is somewhat orthogonal; a \textit{complete} clustering is one where all the data
points that are members of a given class are elements
of the same cluster. The class labels $(0,0,1,1)$ are complete, $C = 1$, with respect to the cluster labels $(1,1,0,0)$, as well as $(0,0,0,0)$. The cluster labels $(0,1,0,1)$ give $C < 1$, since the members of the 1 class are split between the 0 and the 1 cluster.

If the community detection algorithms split one of the target partitions into multiple communities, this will reduce the completeness $C$ but not the homogeneity $H$. $H$ is reduced whenever two nodes from the same target community are assigned by the algorithms to different communities.

The metrics $H$ and $C$ are related by
\begin{equation}
H(t, s) = C(s, t)
\end{equation}
where $t$ is the target or ground truth and $s$ is the partition to be compared. A symmetric linear combination, similar to the $F$ score for precision and recall, known as V-measure,
\begin{equation}
V = \frac{2HC}{H+C}
\end{equation}
can be used to measure the agreement between two label sets when the ground truth is not known. {\color{black} This measure is identical to the commonly used similarity metric `Normalised Mutual Information(NMI)'\footnote{\url{https://scikit-learn.org/stable/modules/generated/sklearn.metrics.v_measure_score.html} (Accessed April 2019)}.  

\begin{figure}
    \centering
    \includegraphics[width=0.5\textwidth]{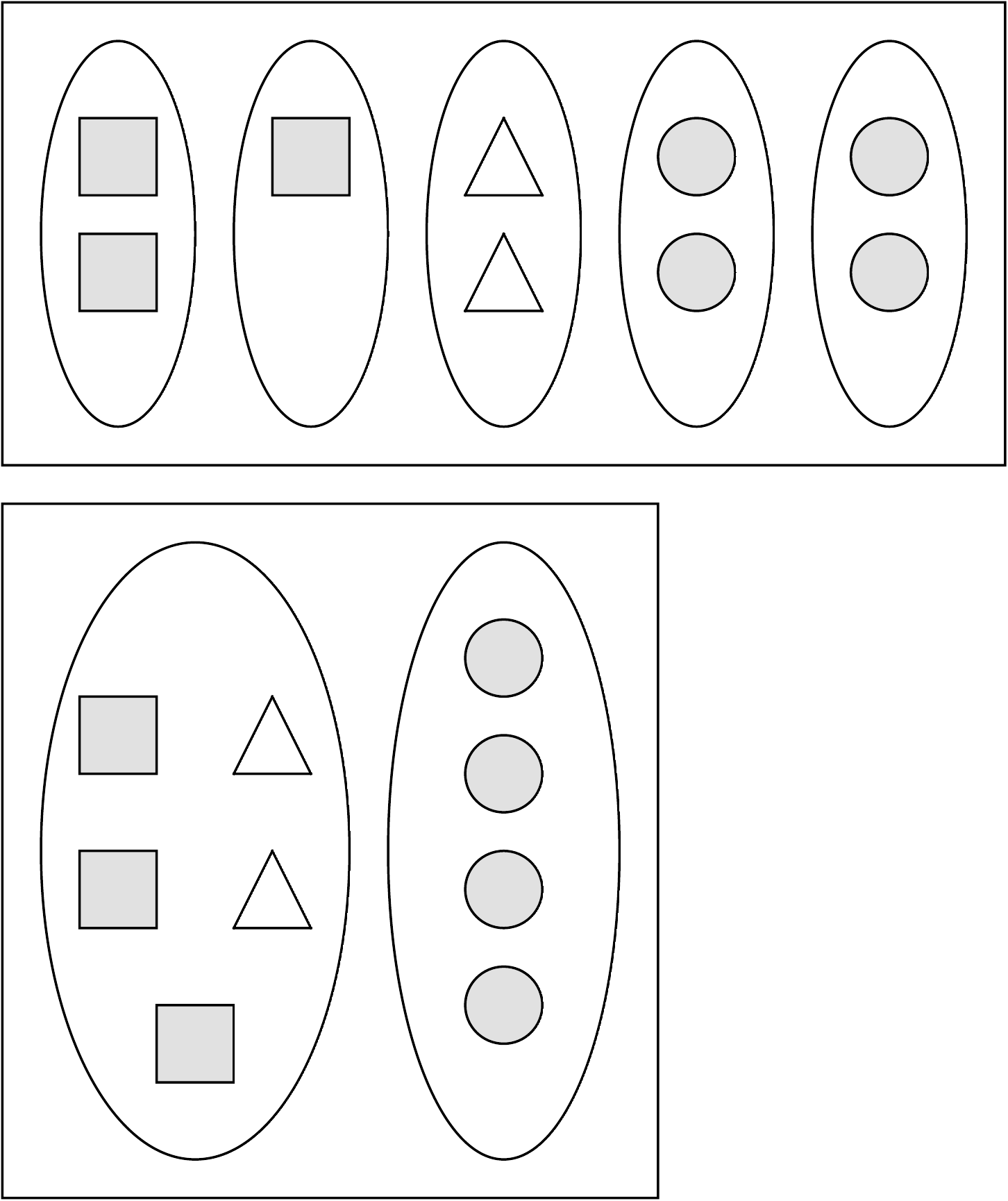}
    \caption{{\color{black}The shapes indicate the ground truth i.e. 3 communities, squares with squares, circles with circles and triangles with triangles. The top partition has $H=1,C=0.67,V=0.8$ while the bottom partition has $H=0.65,C=1,V=0.79$ demonstrating similar V-measure/NMI scores can arise from very different partitions.}}
    \label{fig:completeness}
\end{figure}

It is possible for $H$ and $C$ to be very different, consider the case of assigning all nodes to the same community, this would give $C=1$ and $H \ll 1$. Figure \ref{fig:completeness} shows a simple example of how different partitions can have very different $H$ and $C$ scores but similar V-measure/NMI. Thus we shall examine both $H$ and $C$ as well as V-measure/NMI, so as to better understand the behaviour of these algorithms.}

\section{Results on Synthetic Graphs}\label{sec:syn}

In order to see how the modularity and structure of detected communities depends on the definition of modularity and the projection, we generate a number of synthetic graphs according to the procedure in section \ref{sec:Synthetic}. {\color{black} We will first consider the  case where both top and bottom nodes have a Poisson degree distribution, then the case which arises most in practice, where top and bottom nodes have power-law degree sequences. We will show how modularity, number of detected communities and the target similarity metrics change as a function of $p$ the amount of inter-community coupling. This will allow us to distinguish where the different algorithms and modularity functions perform best.}

We set the target number of left and right nodes in each of the 4 communities to be 250. We generate 100 graph realisations for each set of parameters i.e. the mean degrees $\lambda_1, \lambda_2$ for Poisson distributions and the exponents, $\mu_1, \mu_2$, for power distributions; as well as for every value of the mixing probability $p$. When we project we always project onto the bottom node set, corresponding to the nodes with distributions parameterised by $\lambda_1$ or $\mu_1$. Each of the algorithms is run 10 times on each graph as the final partition found by the Louvain method is sensitive to the starting conditions, the highest modularity partition from each of the 10 runs is selected. Errors are 95\% bootstrap confidence intervals, and are in most cases smaller than the size of the plot symbols.

\subsection{Poisson over Poisson}

\begin{figure}
    \centering
    \includegraphics[width=\textwidth]{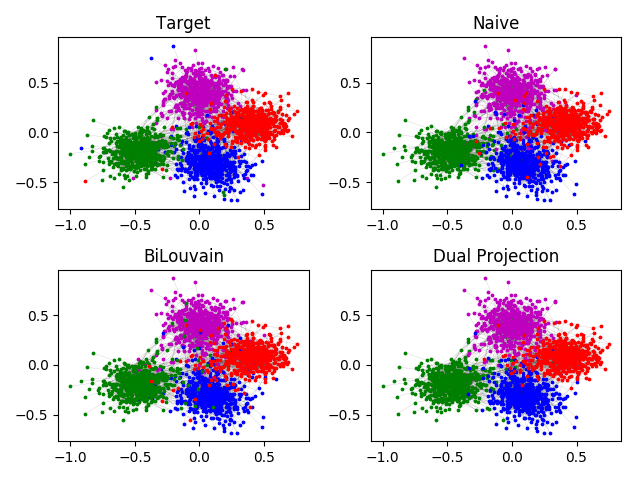}
    \caption{Community detection in a bipartite graph where both node sets have a Poisson degree distribution with mean $5$. The target number of nodes in each community is $N=250$, $M=250$. Mixing parameter $p=0.1$.}
    \label{fig:example}
\end{figure}
We consider the case where the top and bottom nodes both have Poisson degree distributions with mean degree $\lambda_1$ and $\lambda_2$. 
Figure \ref{fig:example} shows the bipartite graph, {\color{black} with the communities detected according algorithms 1, 2 and 3 for a low value of the mixing parameter $p$.} In this case all three methods reproduce the target community structure and agree with the community structure implied by the force directed layout.

\begin{figure}
    \centering
    \includegraphics[width=\textwidth]{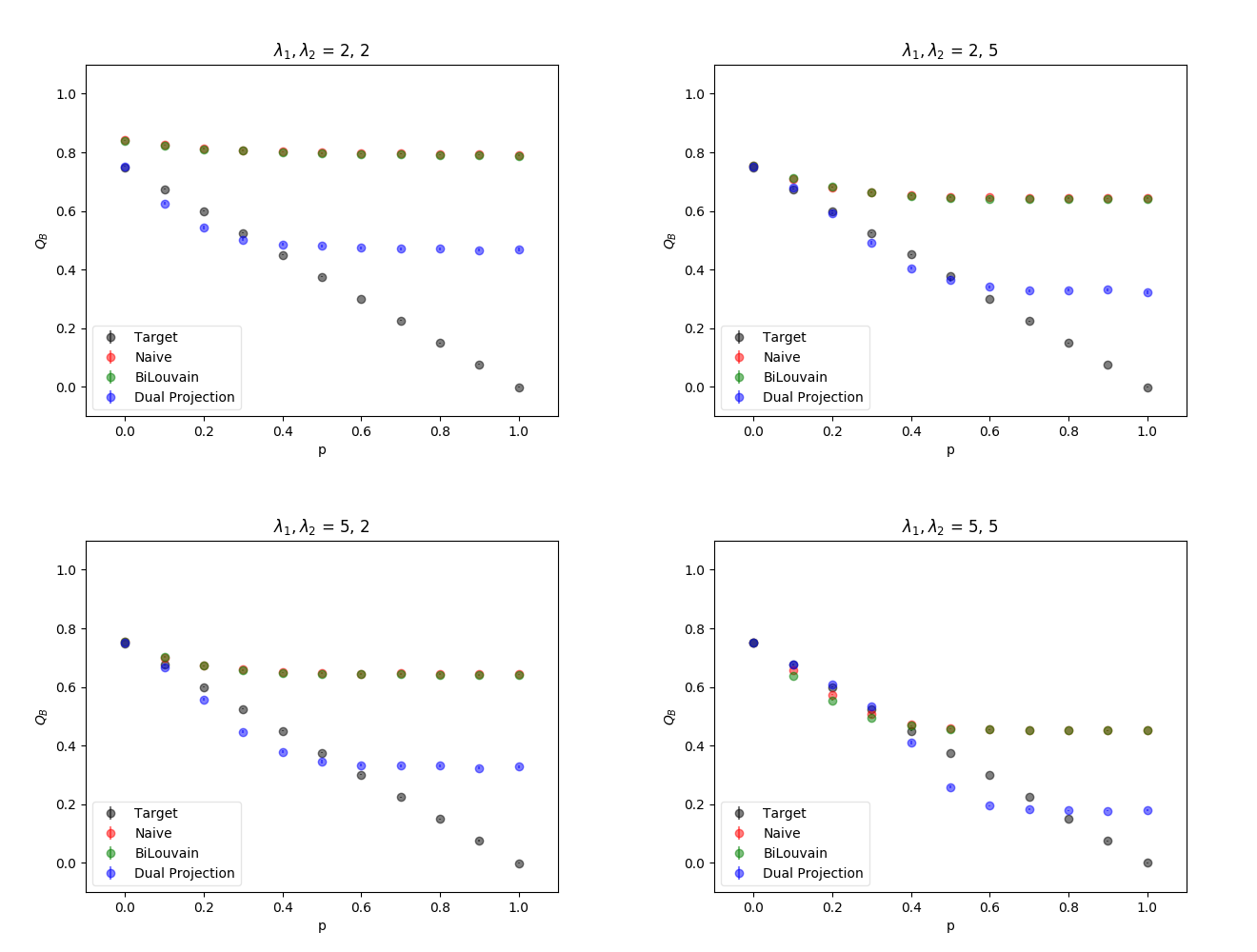}
    \caption{Barber modularity for top and bottom nodes with a Poisson degree distribution as a function of the mixing parameter $p$.}
    \label{fig:poissonbimod}
\end{figure}

Figure \ref{fig:poissonbimod} shows the Barber modularity of the target partition and the maximum modularity partitions found by each of the 3 bipartite algorithms {\color{black}(1,2 and 3)}.  The graph shows that the Naive and BiLouvain algorithms find partitions with essentially the same Barber modularity, where the dual projection approach finds partitions with significantly lower Barber modularity. The modularity of the target partition falls linearly with $p$. {\color{black}This is a demonstration that high modularity should not be taken to imply a good community detection. The target partition, which we hope the modularity maximisation algorithms recover, actually has lower modularity than the detected ones. }

\begin{figure}
    \centering
    \includegraphics[width=\textwidth]{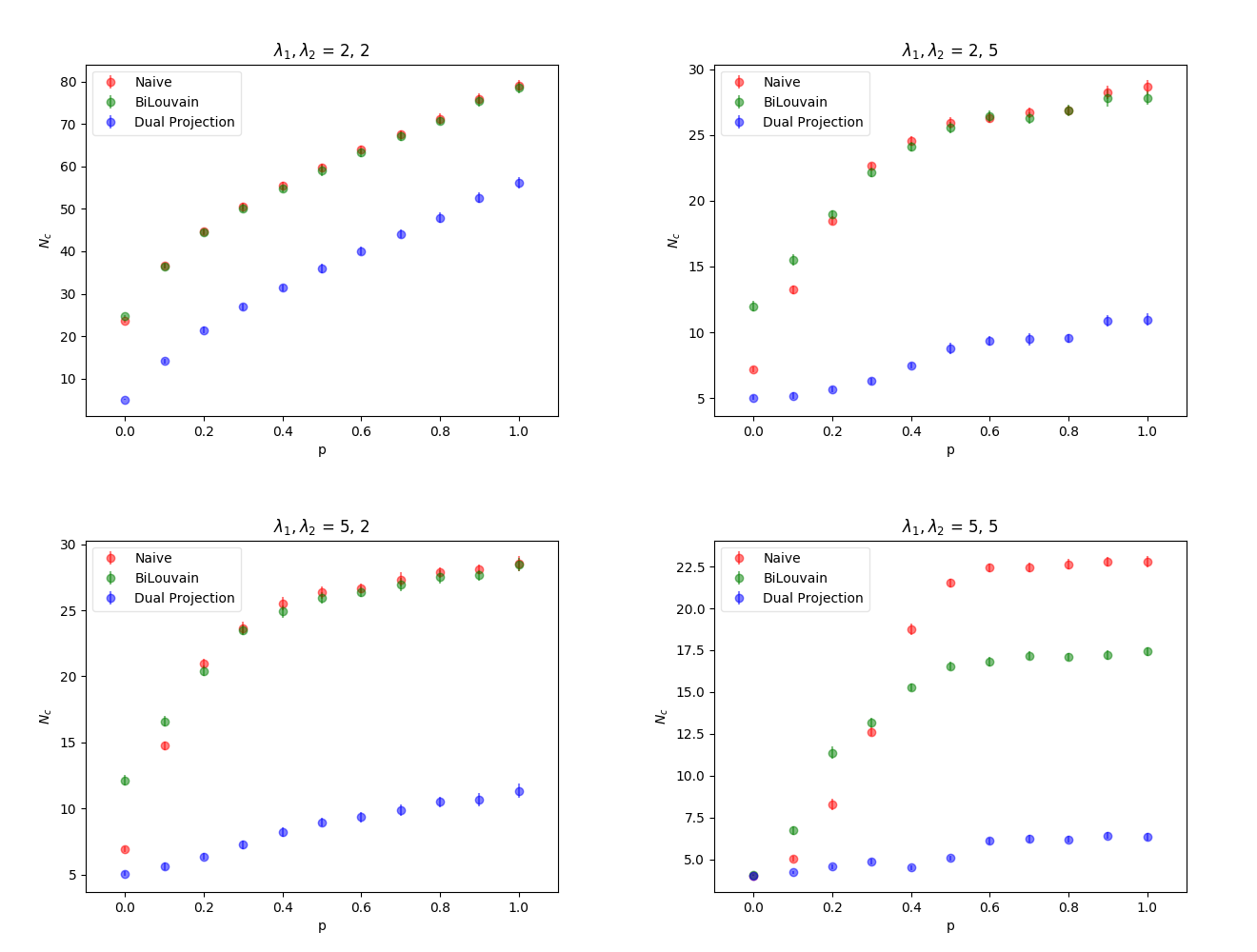}
    \caption{Number of communities found by the bipartite community detection algorithms for graphs with Poisson degree distributions as a function of the mixing parameter $p$.}
    \label{fig:poissonbinumc}
\end{figure}

{\color{black}Figure \ref{fig:poissonbinumc} shows the number of partitions found by each of the bipartite algorithms. The target is 4. Especially for the low degree graph $\lambda_1 = \lambda_2 = 2$ the clustering algorithms detect sub-structure in the target communities resulting in a high number of detected communities. The Dual Projection approach produces the number of communities closest to the target.}

\begin{figure}
    \centering
    \includegraphics[width=\textwidth]{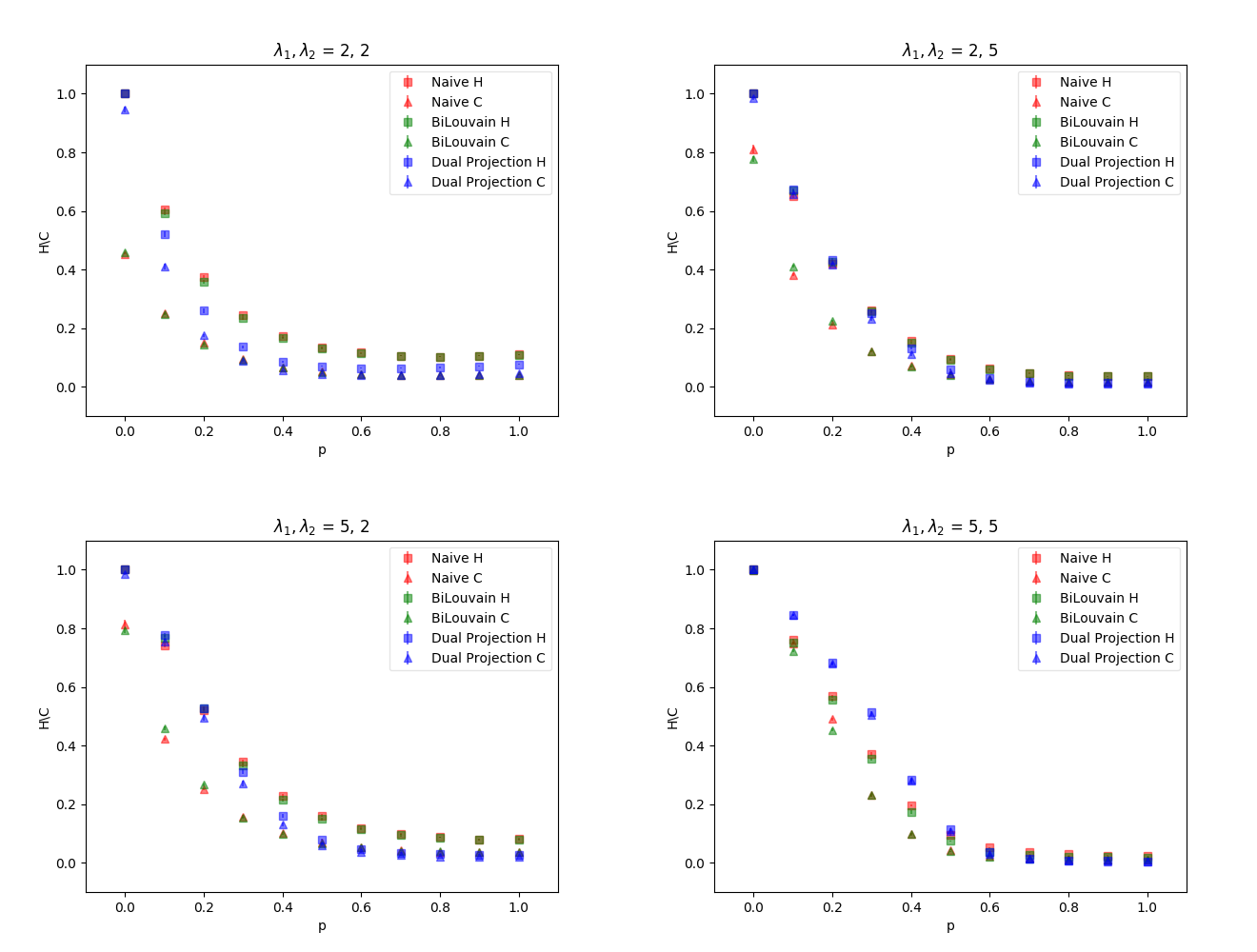}
    \caption{Homogeneity and completeness for communities found by the bipartite community detection algorithms for graphs with Poisson degree distributions as a function of the mixing parameter $p$. Squares are homogeneity values, triangles are completeness values. These algorithms find partitions which are more homogeneous than complete. }
    \label{fig:poissonbihc}
\end{figure}

{\color{black} Figure \ref{fig:poissonbihc} shows that the communities detected by the Naive and BiLouvain methods tend to be more homogeneous than complete, $H>C$. This indicates that these community detection algorithms tend to split large communities into numerous smaller communities.} The Dual Projection approach produces partitions with lower $H$ and larger $C$. This method detects fewer, larger communities which seems to improve the algorithms' ability to resolve the target structure when the network is more densely connected (large $\lambda$).

\subsubsection{Projection}

\begin{figure}[htb]
    \centering
    \includegraphics[width=\textwidth]{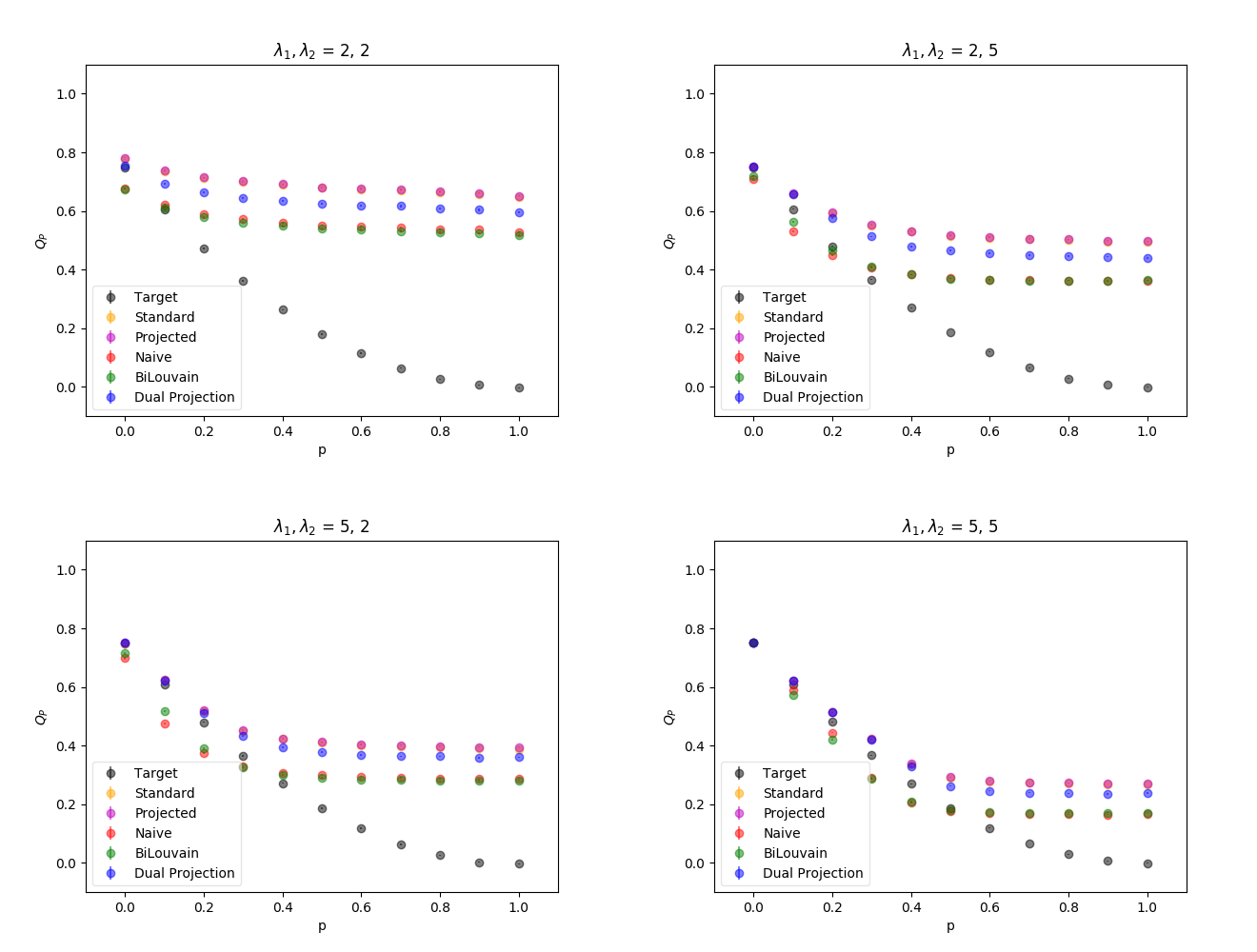}
    \caption{Projected modularity for bottom projection of graphs with a Poisson degree distribution as a function of the mixing parameter $p$. The Projected (magenta) and Standard (orange) data points overlap completely.}
    \label{fig:poissonprojmod}
\end{figure}

\begin{figure}
    \centering
    \includegraphics[width=\textwidth]{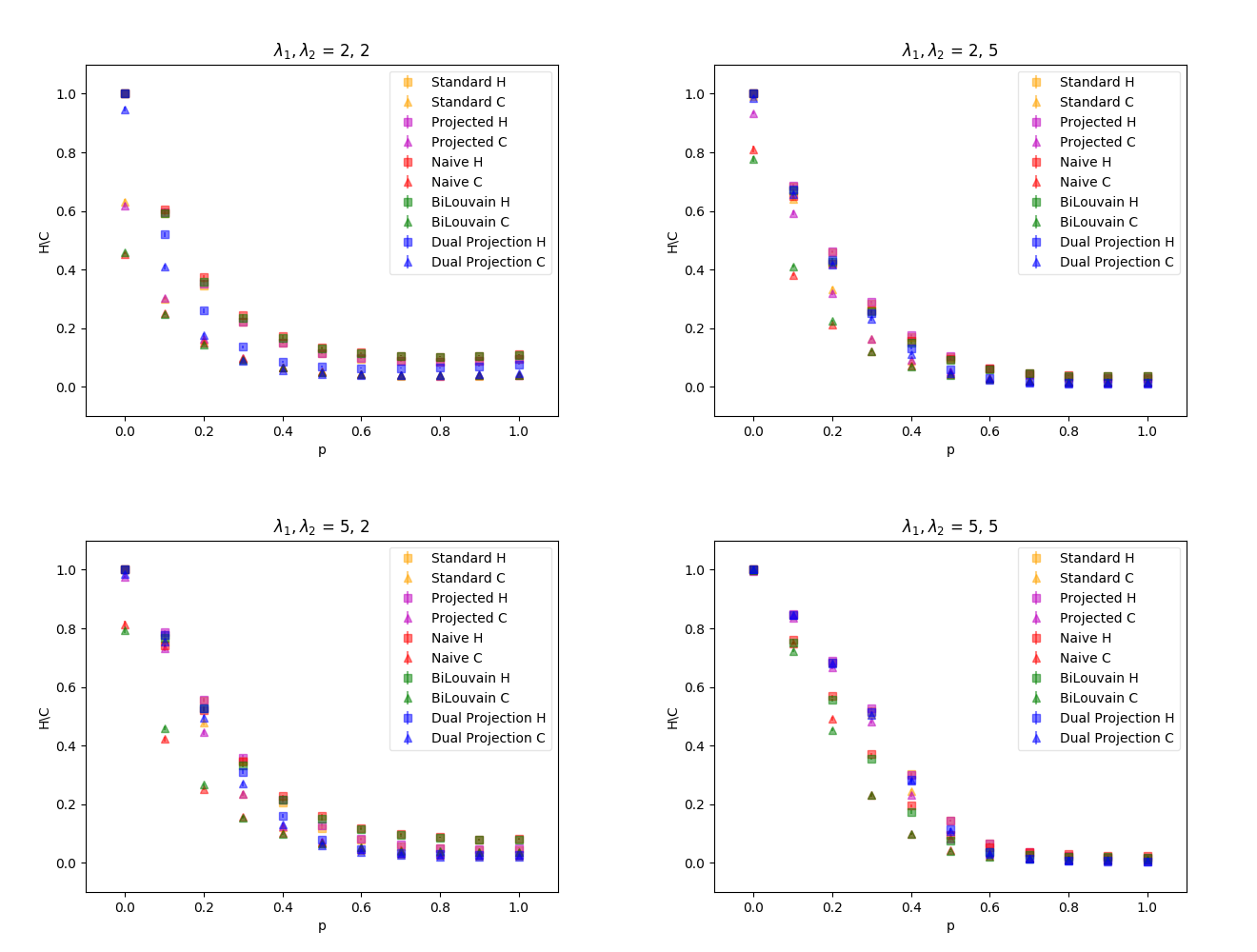}
    \caption{Homogeneity and completeness for communities found by the projected community detection algorithms for graphs with Poisson degree distributions as a function of the mixing parameter $p$. Squares are homogeneity values, triangles are completeness values. As in the bipartite case, we find partitions which are more homogeneous than complete.}
    \label{fig:poissonprojhc}
\end{figure}

We now work with the projected graph. Figure \ref{fig:poissonprojmod}, shows the projected modularity, $Q_P$, as a function of $p$. {\color{black} Similarly to the unprojected case, despite the fact that algorithms 4 and 5 optimise two different modularity functions, the partitions found with both approaches are very similar. The $Q_P$ value for the partitions found by algorithms 4 and 5 are virtually identical. In the case where we find communities on the bipartite graph and then project (methods 1, 2 and 3), the $Q_P$ values are lower, as would be expected since these algorithms optimise a different objective function. The real measure of the partition quality is agreement with the target community structure.

This is measured in Figure \ref{fig:poissonprojhc}, which shows the homogeneity and completeness scores as a function of $p$. Once again we find that the detected partitions are more homogeneous than complete i.e. members of the same community are put in the same partition class, but large communities tend to get split up. The key comparison here is between the algorithms which work on the whole graph (Naive, BiLouvain and Dual Projection) compared to the two projected algorithms (Standard and Projected). The Standard and Projected algorithms are at a similar level of agreement with the target partition as the bipartite algorithms. This will be contrary to some expectations which hold that the projection step masks the true structure of the network. These results imply, for networks with Poisson degree, projection does not destroy community structure.}

\subsection{Power over Power}

\begin{figure}[htp]
    \centering
    \includegraphics[width=\textwidth]{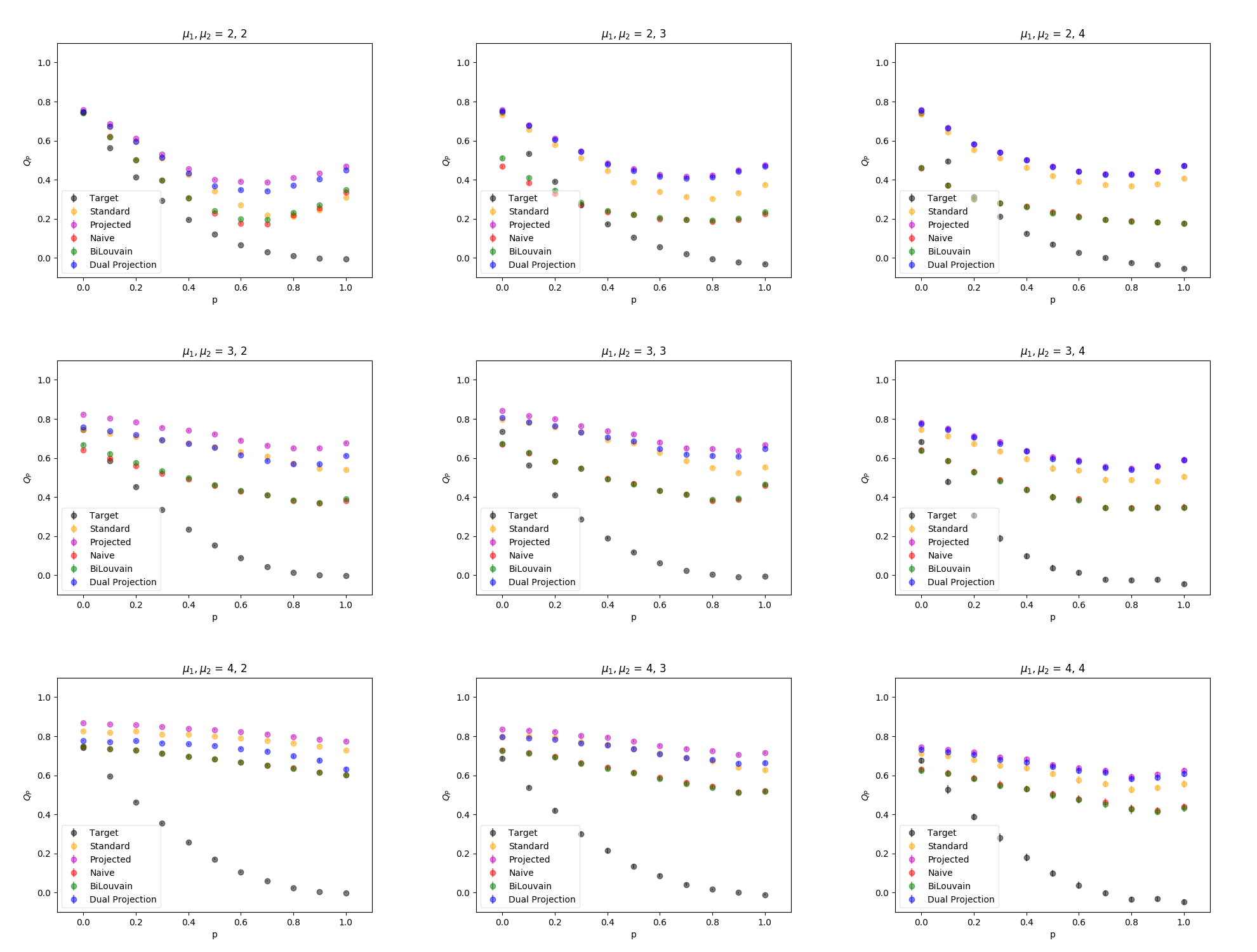}
    \caption{Projected modularity for bottom projection of graphs with a power degree distribution as a function of the mixing parameter $p$}
    \label{fig:powerprojmod}
\end{figure}

\begin{figure}
    \centering
    \includegraphics[width=\textwidth]{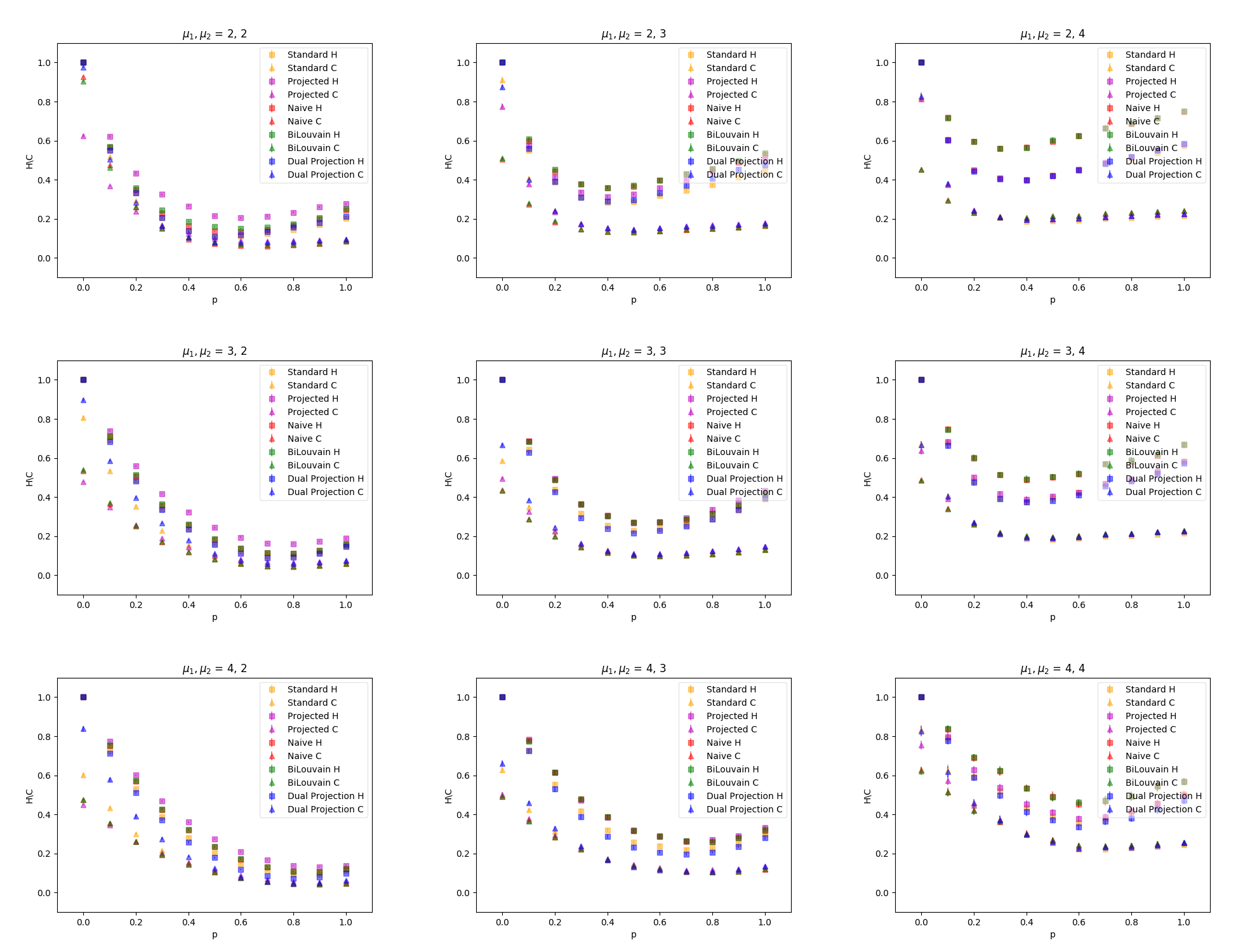}
    \caption{Homogeneity and completeness for communities found by the projected community detection algorithms for graphs with power degree distributions as a function of the mixing parameter $p$.}
    \label{fig:powerprojhc}
\end{figure}

{\color{black} We now turn to the case which arises most frequently in practice, when the degree distributions of both top and bottom nodes follow a power law. Figure \ref{fig:powerprojmod} shows the projected modularity as a function of $p$. Like the Poisson case, as the disorder, $p$, increases the modularity of the target partition decreases while the modularity of the detected partition remains high, regardless of the algorithm or modularity function that is optimised. 

Figure \ref{fig:powerprojhc} shows homogeneity and completeness as a function of $p$. This figure is the key guide to determining a practical heuristic. First, as in the Poisson case, the detected partitions are more homogeneous than complete $H>C$. Thus if two nodes are placed in the same community by the algorithm they are likely to really be in the same community, however the different communities may not really be distinct. The second key observation is that} when the top node set (which is projected away) has a larger magnitude exponent than the bottom node set ($\mu_2 > \mu_1$), the Naive and BiLouvain methods perform much better (in terms of $H$) than the others. When the top nodes have an exponent that is similar or smaller in magnitude ($\mu_2 \leq \mu_1$), the Projected algorithm performs best in terms of homogeneity $H$ and the Dual Projection produces partitions with the highest completeness $C$.

\begin{figure}
    \centering
    \includegraphics[width=\textwidth]{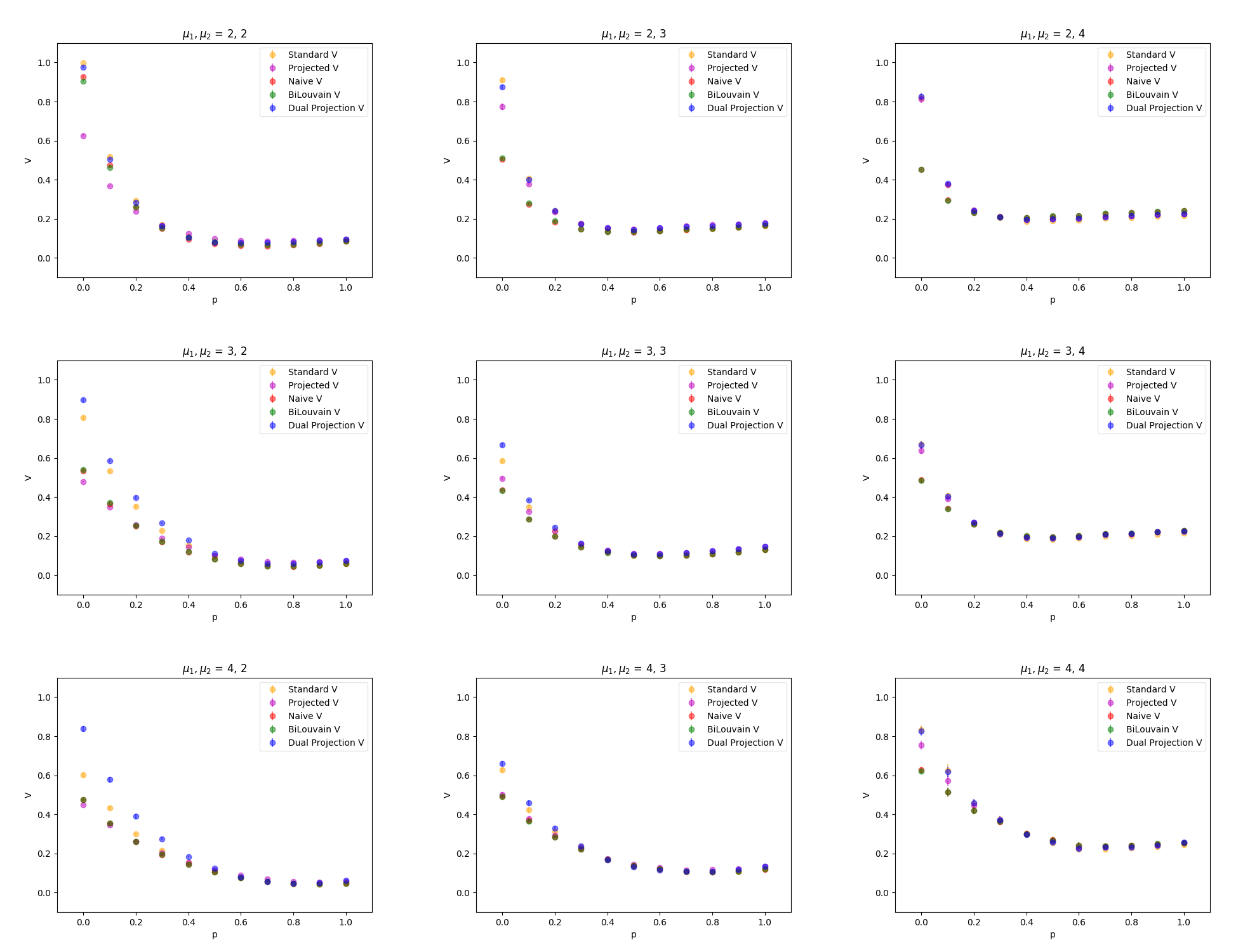}
    \caption{{\color{black}V-measure/NMI for communities found by the projected community detection algorithms for graphs with power degree distributions as a function of the mixing parameter $p$.}}
    \label{fig:powerprojv}
\end{figure}

{\color{black}Figure \ref{fig:powerprojv} simplifies Figure \ref{fig:powerprojhc}. Here we combine homogeneity and completeness scores into the V-measure/NMI score, keeping in mind the information loss this entails. This figure shows that across the range of exponents considered here the Dual Projection algorithm performs best. Although it produces less homogeneous clustering, the higher completeness makes it the method which produces clustering most similar to the target. The V-measure/NMI considers homogeneity and completeness to be equally important. If one of these is more important to us than the other we might want to use a different algorithm. For example, if $H$ is important then, based on Figure \ref{fig:powerprojhc} if $\mu_2 > \mu_1$ the Naive or BiLouvain methods are best, if $\mu_2 < \mu_1$ the Projected method gives largest $H$ and for $\mu_2 = \mu_1$ the projected method seems better for $\mu = 2,3$ and the Naive or BiLouvain better when $\mu_1 = 4$.}

\section{Real Examples}\label{sec:real}

We now examine these methods on four real data sets obtained from the KONECT database \cite{konect}. These are 
\begin{enumerate}
    \item Writers and Works from DBpedia \cite{konect:dbpedia2}. 89356 + 46213 nodes (writers + works). 144340 edges. Projecting onto writers.
    \item Scientific Collaborations from \cite{newman:2001}. 16726 + 22015 vertices (authors + papers). 58595 edges. Projecting onto authors.
    \item Crimes \cite{konect:2016:moreno_crime}. 829 + 551 vertices (people and crimes). 1476 edges. Projecting onto people. 
    \item The famous `Southern Women' social graph (women and events) from \cite{konect:southernwomen}. 18 +14 vertices (women + events). 89 edges. Projecting onto women.
\end{enumerate}

\begin{figure}[htb]
    \centering
    \includegraphics[width=\textwidth]{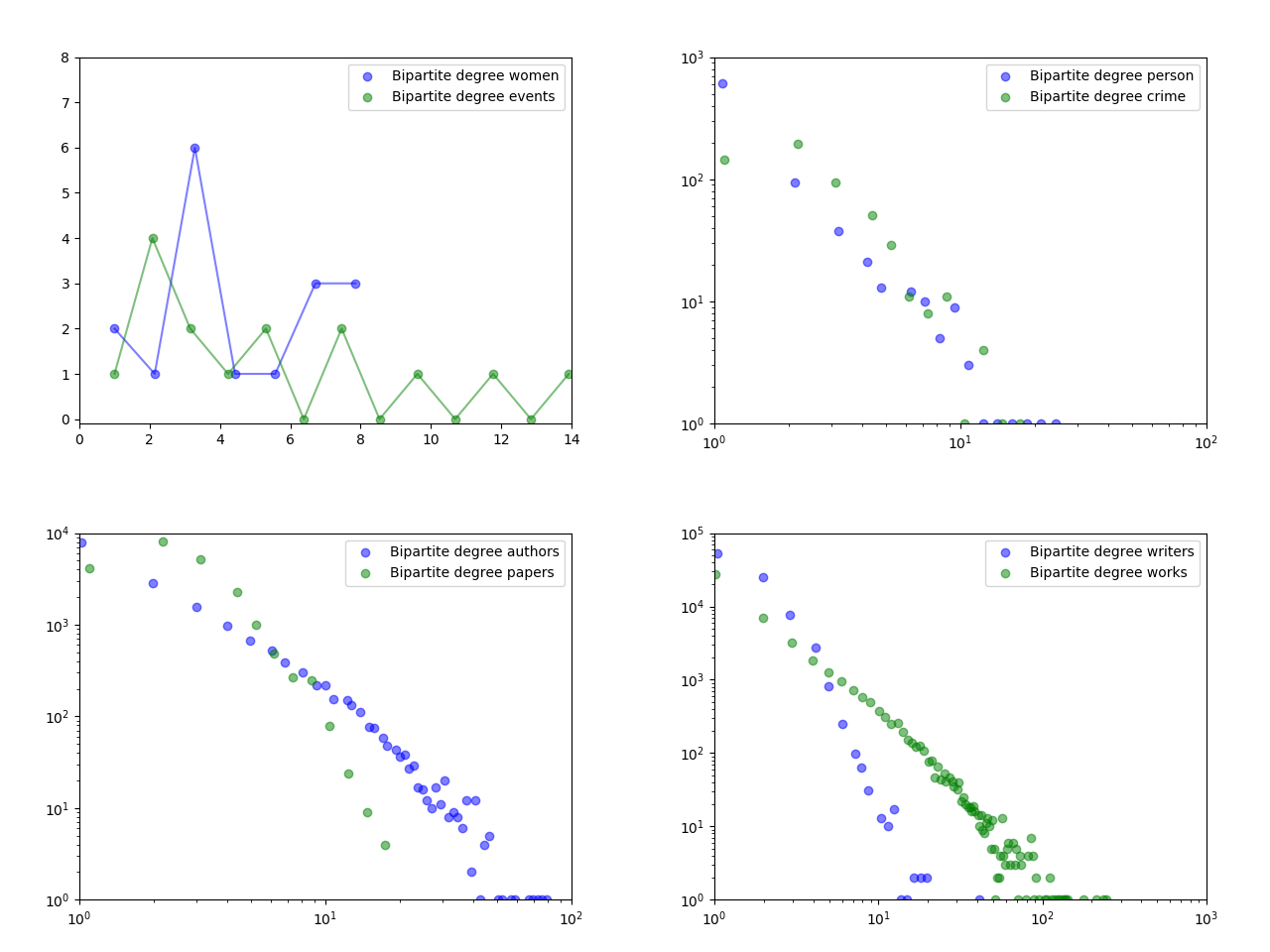}
    \caption{Degree distributions $q$ and $d$ for the 4 real datasets left to right top to bottom: Southern Women, Crime, Collaborations, Writers.}
    \label{fig:realdegree}
\end{figure}

We first examine the degree distributions of each node set in Figure \ref{fig:realdegree}. Apart from the small Southern Women data set, a significant portion of the degree distributions can be reasonably modelled as a power law. It is not the aim of this paper to determine the best fit to the degree distribution (log-normals or other distributions could fit equally well). We simply note that the degree distribution for the Southern Women data doesn't have any obvious simple pattern, while the other three datasets roughly map to: $\mu_1 \sim \mu_2$ for Crime; $\mu_2 > \mu_1$ for Collaborations and $\mu_2 < \mu_1$ for Writers.

\begin{table}
\subfloat[Southern Women]{
\resizebox{0.45\textwidth}{!}{
\begin{tabular}{c|ccccc}
& Standard & Projected & Naive & BiLouvain & Dual Projection \\ \hline
$Q_P$ & 0.152 & 0.152 & 0.089 & 0.070 & 0.152 \\ \hline \hline
H	& Standard & Projected & Naive & BiLouvain & Dual Projection \\ \hline
Standard & 1.000 & 1.000 & 0.728 & 0.846 & 1.000 \\
Projected & 1.000 & 1.000 & 0.728 & 0.846 & 1.000 \\
Naive & 0.459 & 0.459 & 1.000 & 0.859 & 0.459 \\
BiLouvain & 0.429 & 0.429 & 0.690 & 1.000 & 0.429 \\
Dual Projection & 1.000 & 1.000 & 0.728 & 0.846 & 1.000 \\
\hline  \hline
V	& Standard & Projected & Naive & BiLouvain & Dual Projection \\ \hline
Standard & 1.000 & 1.000 & 0.563 & 0.569 & 1.000 \\
Projected & 1.000 & 1.000 & 0.563 & 0.569 & 1.000 \\
Naive & 0.563 & 0.563 & 1.000 & 0.765 & 0.563 \\
BiLouvain & 0.569 & 0.569 & 0.765 & 1.000 & 0.569 \\
Dual Projection & 1.000 & 1.000 & 0.563 & 0.569 & 1.000 \\
\end{tabular}}}
\quad
\subfloat[Crime]{
\resizebox{0.45\textwidth}{!}{
\begin{tabular}{c|ccccc}
& Standard & Projected & Naive & BiLouvain & Dual Projection \\ \hline
$Q_P$ & 0.887 & 0.902 & 0.857 & 0.851 & 0.879 \\ \hline \hline
H	& Standard & Projected & Naive & BiLouvain & Dual Projection \\ \hline
Standard & 1.000 & 0.940 & 0.916 & 0.913 & 0.821 \\
Projected & 0.919 & 1.000 & 0.908 & 0.906 & 0.839 \\
Naive & 0.861 & 0.873 & 1.000 & 0.966 & 0.763 \\
BiLouvain & 0.840 & 0.852 & 0.946 & 1.000 & 0.729 \\
Dual Projection & 0.958 & 1.000 & 0.947 & 0.923 & 1.000 \\
\hline  \hline
V	& Standard & Projected & Naive & BiLouvain & Dual Projection \\ \hline
Standard & 1.000 & 0.929 & 0.887 & 0.875 & 0.884 \\
Projected & 0.929 & 1.000 & 0.890 & 0.878 & 0.912 \\
Naive & 0.887 & 0.890 & 1.000 & 0.956 & 0.845 \\
BiLouvain & 0.875 & 0.878 & 0.956 & 1.000 & 0.815 \\
Dual Projection & 0.884 & 0.912 & 0.845 & 0.815 & 1.000 \\
\end{tabular}}}\\
\subfloat[Collaboration]{
\resizebox{0.45	\textwidth}{!}{\begin{tabular}{c|ccccc}
& Standard & Projected & Naive & BiLouvain & Dual Projection \\ \hline
$Q_P$ & 0.877 & 0.883 & 0.857 & 0.860 & 0.848 \\ \hline \hline
H	& Standard & Projected & Naive & BiLouvain & Dual Projection \\ \hline
Standard & 1.000 & 0.873 & 0.827 & 0.803 & 0.660 \\
Projected & 0.871 & 1.000 & 0.828 & 0.820 & 0.672 \\
Naive & 0.759 & 0.762 & 1.000 & 0.835 & 0.570 \\
BiLouvain & 0.750 & 0.768 & 0.850 & 1.000 & 0.572 \\
Dual Projection & 0.871 & 0.888 & 0.818 & 0.807 & 1.000 \\
\hline  \hline
V	& Standard & Projected & Naive & BiLouvain & Dual Projection \\ \hline
Standard & 1.000 & 0.872 & 0.792 & 0.776 & 0.751 \\
Projected & 0.872 & 1.000 & 0.794 & 0.793 & 0.765 \\
Naive & 0.792 & 0.794 & 1.000 & 0.843 & 0.672 \\
BiLouvain & 0.776 & 0.793 & 0.843 & 1.000 & 0.670 \\
Dual Projection & 0.751 & 0.765 & 0.672 & 0.670 & 1.000 \\
\end{tabular}}}
\quad
\subfloat[Writers]{
\resizebox{0.45\textwidth}{!}{\begin{tabular}{c|ccccc}
& Standard & Projected & Naive & BiLouvain & Dual Projection \\ \hline
$Q_P$ & 0.940 & 0.948 & 0.890 & 0.906 & 0.918 \\ \hline \hline
H	& Standard & Projected & Naive & BiLouvain & Dual Projection \\ \hline
Standard & 1.000 & 0.941 & 0.905 & 0.905 & 0.815 \\
Projected & 0.929 & 1.000 & 0.897 & 0.903 & 0.815 \\
Naive & 0.884 & 0.888 & 1.000 & 0.919 & 0.783 \\
BiLouvain & 0.887 & 0.898 & 0.923 & 1.000 & 0.791 \\
Dual Projection & 0.963 & 0.976 & 0.947 & 0.953 & 1.000 \\
\hline  \hline
V	& Standard & Projected & Naive & BiLouvain & Dual Projection \\ \hline
Standard & 1.000 & 0.935 & 0.895 & 0.896 & 0.883 \\
Projected & 0.935 & 1.000 & 0.893 & 0.901 & 0.889 \\
Naive & 0.895 & 0.893 & 1.000 & 0.921 & 0.857 \\
BiLouvain & 0.896 & 0.901 & 0.921 & 1.000 & 0.865 \\
Dual Projection & 0.883 & 0.889 & 0.857 & 0.865 & 1.000 \\
\end{tabular}}}
\caption{Projected Modularity, Homogeneity and V-measure for the partitions found by each of the 5 algorithms.}\label{tab:metrics}
\end{table}

We perform community detection using all 5 different algorithms. Table \ref{tab:metrics} shows the homogeneity and completeness scores of every algorithm against every other algorithm (remembering that $H(t,s) = C(s,t)$). There is no `ground truth' in this case so we use the (symmetric) V-measure to measure the similarity of the detected partitions.

Broadly, we can say the Standard and Projected methods give results with similar values of $Q_P$, with the Projected method giving the highest value (as it should since this is what it is attempting to maximise!). Next is the Dual Projection approach (except for the Collaboration network) and then the Naive and BiLouvain methods. In terms of the partition similarity, the Standard and Projected methods give the most similar partitions, the Naive and BiLouvain methods also have relatively high similarity. The Dual Projection approach gives the least similar partitions, especially for the Collaboration graph. 

\section{Conclusions}\label{sec:conclusion}

We have explored the problem of community detection on real and synthetic bipartite graphs. {\color{black} We have constructed a modularity metric $Q_P$ which is more appropriate for community detection on projected bipartite graphs than the standard modularity. We then performed numerous experiments attempting to optimise different modularity functions.}

Generally, when applying community detection algorithms in practice, the goal is not simply to find clusters of densely inter-connected nodes, but to recover some aspect of the generative process which produced the graph. We expect that nodes in the same communities will have some characteristic which results in higher intra-community connectivity and hence higher modularity e.g. acting in the same language or writing in the same genre. Modularity is a proxy measure and modularity maximisation simply finds communities with more internal than external connections than would be expected in the null model. 

Our results on synthethic graphs show that modularity maximisation produces communities which are more homogeneous than complete with respect to the partitions defined by the generative process by which we created our networks. These algorithms tend to split potential communities more readily than joining them. {\color{black} The Dual Projection method seems to be best at balancing $H$ and $C$ and giving the largest NMI, while the other algorithms optimise $H$}. Tuning the resolution parameter, \cite{Fortunato36, lambiotte2008laplacian}, could also be an effective way to address this. Table \ref{tab:metrics} shows that while all algorithms find very high modularity partitions ($Q_P > 0.8$), these high modularity partitions can still be quite dissimilar, again underlining the need for caution when using modularity to determine quality of the partitioning.

To answer our fundamental question - is community detection on the projected graph sufficient to recover the true community structure - the key results are figure \ref{fig:powerprojhc} and \ref{fig:powerprojv}. When the exponent of the nodes to be projected out is larger, $\mu_2 > \mu_1$, then the Naive and BiLouvain algorithms find {\color{black} the most homogeneous partitions}. In this case the top node set will be dominated by a small number of high degree nodes. The projection will result in cliques which hide the real structure. When $\mu_2 < \mu_1$ the Standard and Projected algorithms produce the {\color{black} the most homogeneous partitions}. Here the high degree nodes are more likely to be in the bottom set, so cliques are not formed as readily when projecting and communities detected on the projected network agree with the target communities. 

Ultimately, our advice for community detection on bipartite graphs is:
\begin{itemize}
    \item {\color{black} Use the Dual Projection approach.
    \item If this is not possible because the agglomeration step is too time consuming, the necessary software is not available or, if homogeneity is the most important consideration then:}
    \begin{itemize}
        \item Measure degree distributions of the top and bottom node sets.
    \item If $\mu_2 > \mu_1$ use the bipartite graph and optimise $Q_B$ if possible, otherwise optimising $Q$ will produce a similar partition.
    \item If $\mu_2 \leq \mu_1$ project the network and optimise $Q_P$ if possible, otherwise optimising $Q$ will produce a similar partition.
    \end{itemize}
\end{itemize}
If the top node set is not available then considerations about whether its degree distribution would be heavier or lighter tailed, or if it would have many more or fewer nodes will give some guidance on if the detected communities can be trusted. The observation that optimising $Q$ gives partitions which are almost as good as the ones obtained by optimising $Q_B$ and $Q_P$ (in terms of similarity to the target and modularity) implies that even if the bipartite structure is unavailable we can still do a good job of detecting communities. 

Finally it must always be acknowledged that simply finding a high modularity partition is not evidence of community structure. If there is underlying community structure, modularity maximisation is likely to detect only a certain aspect of it. Our results on homogeneity and completeness suggest nodes from the same target community are likely to be in the same partition, but detected partitions are also likely biased to be smaller and more numerous than the `real' communities. One should compare the results of different community detection methods e.g. \cite{holland1983stochastic,newman2001scientific,newman2013spectral} as well as methods which look for other structural features of the network e.g. core-periphery structure \cite{kojaku2017finding}. Most important however, are sanity checks based on the nature of the nodes: if a community detection algorithm puts all Italian actors in one community and all French actors in another, it is probably working as intended, if not, one needs to take a serious look at the method.

\section*{Acknowledgements}
I would like to thank Hywel Williams, Iain Weaver and Tristan Cann for inspiring me to think about this problem and for useful discussions.

\section*{References}
\bibliography{mybibfile}

\end{document}